\documentclass[12pt,a4paper,onecolumn, prb]{article}
\usepackage[utf8]{inputenc}
\usepackage[T1]{fontenc}
\usepackage{amsmath}
\usepackage{amsfonts}
\usepackage{amssymb}
\usepackage{graphicx}
\usepackage{authblk}
\usepackage{cite}
\usepackage[colorlinks=true, linkcolor=blue, citecolor = blue]{hyperref}
 \usepackage[margin=2.5cm]{geometry}

\author[1,2]{Antonija Grubi\v{s}i\'{c}-\v{C}abo \thanks{a.grubisic-cabo@rug.nl,}}
\author[1,3]{Jimmy C. Kotsakidis}
\author[4,5]{Yuefeng Yin}
\author[5,6,7]{Anton Tadich}
\author[1]{Matthew Haldon}
\author[1]{Sean Solari}
\author[7]{John Riley}
\author[7]{Eric Huwald}
\author[8]{Kevin M. Daniels}
\author[9]{Rachael L. Myers-Ward}
\author[1,5]{Mark T. Edmonds}
\author[4,5]{Nikhil Medhekar}
\author[10]{D. Kurt Gaskill}
\author[1,5]{Michael S. Fuhrer \thanks{michael.fuhrer@monash.edu}}

\date{}

\affil[1]{School of Physics and Astronomy, Monash University, Clayton, VIC 3800, Australia}
\affil[2]{Zernike Institute for Advanced Materials, University of Groningen, 9747 AG Groningen, The Netherlands}
\affil[3]{Laboratory for Physical Sciences, University of Maryland, College Park, MD, 20740, USA}	
\affil[4]{Department of Materials Science and Engineering, Monash University, Clayton, VIC 3800, Australia}	
\affil[5]{ARC Centre of Excellence in Future Low Energy Electronics Technologies (FLEET), Monash University, Clayton, VIC 3800, Australia}
\affil[6]{Australian Synchrotron, Clayton, VIC 3168, Australia}
\affil[7]{Department of Physics, La Trobe University, Bundoora, VIC 3086, Australia}
\affil[8]{Department of ECE, University of Maryland, College Park, MD 20742, USA}
\affil[9]{U.S. Naval Research Laboratory, Washington, D.C., 20375, USA}
\affil[10]{Institute for Research in Electronics and Applied Physics, University of Maryland, College Park, MD 20742, USA}

\title{Quasi-free-standing AA-stacked bilayer graphene induced by calcium intercalation of the graphene-silicon carbide interface}

\begin{document}
	\maketitle
	
	\clearpage
	\begin{abstract}

We study quasi-freestanding bilayer graphene on silicon carbide intercalated by calcium. The intercalation, and subsequent changes to the system, were investigated by low-energy electron diffraction, angle-resolved photoemission spectroscopy (ARPES) and density-functional theory (DFT). Calcium is found to intercalate only at the graphene-SiC interface, completely displacing the hydrogen terminating SiC. As a consequence, the system becomes highly n-doped. Comparison to DFT calculations shows that the band dispersion, as determined by ARPES, deviates from the band structure expected for Bernal-stacked bilayer graphene. Instead, the electronic structure closely matches AA-stacked bilayer graphene on Ca-terminated SiC, indicating a spontaneous transition from AB- to AA-stacked bilayer graphene following calcium intercalation of the underlying graphene-SiC interface.

	\end{abstract}
	\section{Introduction}

Graphene, a single layer of graphite \cite{Geim_Novoselov_Graphene1}  is notable for its unique bandstructure with massless Dirac Fermions \cite{Geim_Novoselov_MasslessFermions}, which give rise to a plethora of exotic physical phenomena, such as a $\pi$-Berry phase \cite{GrapheneBerryPhase_PKim, GrapheneBerryPhase_ARPES_Liu2011, Graphene_BerryPhase_SteveLouieARPES2011}, Klein tunnelling \cite{Graphene_KleinTunnelig_Geim2006} and an unusual quantum Hall effect \cite{GrapheneBerryPhase_PKim}.

In contrast, the most typical form of bilayer graphene, so called AB- or Bernal stacked bilayer graphene (Supplementary Information Figure S3\textbf{A, B}), has a completely different electronic structure with massive, yet gapless, Dirac fermions, and a Berry phase of 2$\pi$ \cite{BL_Gr_ARPES_EliScience,GrapheneToGraphite_TB, McCann_Koshino_BLgrTB,BL_Graphene_BerryPhasE_Geim}.
In principle, other types of stacking, such as AA-stacking (Supplementary Information Figure S3\textbf{C, D}), exist. AA-stacking is a metastable stacking, where graphene layers lie directly above one another. Consequently, AA-stacked graphene has an electronic structure which can be considered as a superposition of two single-layer spectra, preserving massless Dirac fermions and a $\pi$-Berry phase \cite{AA_BL_Graphene_TEM, AABLgr_TB_Nori2016}. Despite many interesting properties predicted for AA-stacked bilayer graphene, including a recent prediction that it might host a fractional metal state \cite{AAgraphene_fractionalMetal_PRL2021},  there are very few experimental realisations  \cite{Twisted_AABlg_ELi_NatMat2013, AA_BL_Graphene_TEM, Li_intercal_EMLG_SiC_Johansson_2016, Li_intercal_BLgr_AA_AB_mix, GrSiC_LeidenLEEM2022,GrSiC_Solitons_AA}. Out of the few reported cases, the majority have been found in lithium intercalated systems \cite{Li_intercal_EMLG_SiC_Johansson_2016,Li_intercal_BLgr_AA_AB_mix}, or contained within very small regions otherwise surrounded by AB-stacked graphene \cite{GrSiC_Solitons_AA,GrSiC_LeidenLEEM2022}.

One of the most promising methods for graphene production in terms of scalability is the growth of graphene on silicon carbide (SiC) which allows formation of large-scale graphene with high carrier mobility \cite{Starke_and_Riedl_EpiGrapehneSiC2009,Kurt_EMLG_growth,Seyller_Tegenkamps_Schumacher_EpiGrapheneForElectronicsComeback2016, Seyller_WaferScale_GrSiC2009}.
Graphene on SiC can either be epitaxial, i.e. directly grown on the SiC, with a buffer layer in between the graphene and the SiC interface, or quasi-freestanding graphene--most commonly created via hydrogen intercalation of epitaxial graphene \cite{Kurt_QFSgr_growth, Rield_HGR_1stHintercal_PRL2009}, in which graphene retains the properties expected for the isolated layer \cite{H_Gr_SiC_idealGraphene_Starke_PRL2015}. Hydrogen is not the only element that can be used to create quasi-freestanding graphene on SiC by means of intercalation \cite{Gr_SiC_Intercal_MiniReviewBriggs2019}; various other elements can be used, such as gold \cite{Au_Buffer_SiC_pnJunct_KoreaJour, Au_gr_SiC_Rashba_Seyller2016},  iron \cite{Fe_gr_SiC_Intercal_Chung2014, Fe_Gr_SiC_Intercal_Justin2018}, oxygen \cite{O_Gr_SiC_Intercal_Seyller2013}, lithium \cite{Li_Intercal_GrSiO2_Michael, Li_intercal_EMLG_SiC_Johansson_2016, Li_intercal_BLgr_AA_AB_mix, Li_intercal_Swe1}, magnesium \cite{JimmyXPS, Antonija_MgGr_ARPEStoroid, Jimmy_Mg_Gr_Mesas}, calcium \cite{JimmyXPS, CaC6_AnisotropicElPhonCoupling_Valla_PRL2009, CaC6_GammaARPES_ZXShen_NatComm2014, Ca_Intercal_BLgr_SiC_Endo_2020, Ca_Intercal_SiC_CaLi_HasegawaNew, CaIntercal_BLGr_CaC6_Transport_supercond_Shuji_ACSNano2016, Ca_Intercal_SiC_CaLi_HasegawaNew}, antimony \cite{Sn_gr_SiC_Intercal_Seyller2019} and ytterbium \cite{Yb_Intercal_Gr_Oxidation}. The majority of the intercalation studies have been done on epitaxial monolayer and bilayer graphene on SiC, with very few intercalation studies on already quasi-freestanding, hydrogen intercalated, graphene \cite{Li_H_Gr_SiC_Linkopping_SurfSci2012, QFSMLG_SnGe_intercal2019, JimmyXPS}. Of particular interest to us is calcium intercalated graphene, whose study was inspired by the bulk superconducting graphite intercalation compound CaC$_6$ \cite{Ca_Intercal_Graphite_supercond_Sato2009, CaC6_GammaARPES_ZXShen_NatComm2014, Ca_Graphite_11p5K_Nature,Ca_intercal_Graphite11p5K_PRL}. The majority of calcium intercalation experiments have been performed on graphene grown on SiC, as this allows for growth of large-area graphene that can be characterised with various surface characterisation techniques, such as X-ray photoelectron spectroscopy, angle-resolved photoemission spectroscopy (ARPES), low-energy electron diffraction (LEED) and scanning tunnelling microscopy \cite{JimmyXPS, BL_Gr_ARPES_EliScience, C6Ca_CaIntercal_GrapheneThinnestCaC6_Kanetani_PNAS2012, Ca_Gr_vanHove_Eli}. Calcium intercalation is known to strongly n-type dope graphene, an effect which has been extensively studied \cite{BL_Gr_ARPES_EliScience, Ca_Gr_vanHove_Eli}, however, the impact of calcium intercalation on the structural aspects of graphene and the precise positioning of calcium atoms remained somewhat ambiguous \cite{JimmyXPS, CaIntercal_BLGr_CaC6_Transport_supercond_Shuji_ACSNano2016, C6Ca_CaIntercal_GrapheneThinnestCaC6_Kanetani_PNAS2012}. Recent research using X-ray photoelectron spectroscopy (XPS) by Kotsakidis et al. \cite{JimmyXPS} has shed light on this, revealing that calcium is situated at the interface between the SiC substrate and graphene buffer layer, with work by Toyama et al. \cite{Ca_Intercal_SiC_CaLi_HasegawaNew} further confirming that calcium prefers to go to the SiC interface.

In this paper, we report calcium intercalation of quasi-freestanding bilayer graphene (QFSBLG) on SiC. Using a combination of LEED, ARPES and density-functional theory (DFT), calcium is found to intercalate only at the interface between graphene and SiC, fully replacing hydrogen in the structure, and not between the graphene layers. This results in highly n-doped, quasi-freestanding bilayer graphene (Ca-QFSBLG) with a drastically altered electronic structure, as seen by ARPES. Comparison with DFT shows the structure to be in close agreement with AA-stacked bilayer graphene, indicating a spontaneous transition from AB- to AA-stacking, which has not been previously observed for calcium intercalated graphene. % This is akin to the case of lithium intercalated epitaxial bilayer graphene on SiC \cite{Li_intercal_EMLG_SiC_Johansson_2016, Li_intercal_BLgr_AA_AB_mix, DFT_Alkali_EarthAlkali_BLGR_Intercal, Li_H_Gr_SiC_Linkopping_SurfSci2012}, but it was not previously observed by ARPES in the case of calcium intercalation.
	
	\section{Materials and Methods}
	
	\subsection*{Sample preparation}
	QFSBLG samples on SiC were grown on semi-insulating 6H-SiC(0001) substrate as described in Ref. \cite{Kurt_QFSgr_growth}.  
	Sample preparation, ARPES and LEED measurements were carried out at the Toroidal Analyzer endstation at the Soft X-ray Beamline of the Australian Synchrotron. Samples were introduced to ultra-high vacuum (UHV, base pressure of 1 $\times$ 10$^{-10}$ mbar), and annealed over night at 773 -- 823 K. Sample cleanliness was confirmed by LEED and ARPES. A calcium effusion cell was baked at 423 K overnight and outgassed at 588 K. Once the pressure reached 1 × 10$^{-8}$ mbar, the effusion cell was inserted into the UHV preparation chamber. Calcium (dendritic pieces, 99.99\%, Sigma-Aldrich) was intercalated under graphene following modified recipe from Ref. \cite{JimmyXPS}: Calcium was evaporated for 15 min, with the calcium cell held at 688 K, and deposited on the graphene/SiC substrate held at room temperature. The thickness of deposited calcium layer was 22 $\mathring{A}$, as determined by a quartz crystal microbalance. Following the deposition, the graphene/SiC substrate was annealed at 773 K for 15 minutes, in order to facilitate calcium intercalation. \\
	
\subsection*{Angle-resolved photoemission spectroscopy \& Low-energy electron diffraction}

Structural characterisation of samples was undertaken using a LEED (OCITM 3 grid reverse view optics, 200 $\mu$m spot size) at room temperature, in the endstation used for ARPES. ARPES measurements used a toroidal-type angle-resolving endstation \cite{ToroidalMarkII} at the Soft X-Ray Beamline of the Australian Synchrotron. All ARPES data was taken at room temperature with photon energy (h$\nu$) of 100 eV using linearly polarised light at normal incidence to the sample. The beam spot size was 100 $\mu$m × 60 $\mu$m. The binding energy (E$_{Bin}$) scale for all spectra is referenced to the Fermi energy (E$_F$), determined using the Fermi edge of a gold foil reference sample in electrical contact with the sample. The toroidal analyser permits all polar ($\Theta$) emission angles (-90$^\circ$ to +90$^\circ$) to be measured along a high-symmetry azimuth ($\phi$) of the surface containing the $\bar{\Gamma}$ point. This unique geometry allows for measurement of the Dirac cone along the  $\bar{K}-\bar{\Gamma}-\bar{ K}$ high-symmetry direction without the need for complex alignment of the spectrometer. Under this geometry, the polarisation  vector of the X-rays is entirely contained in the detection plane. The estimated momentum and energy resolution are $\approx$ 0.02 $\mathring{A}^{-1}$ and $\approx$ 150 meV.\\ 
	
	\subsection*{Density-functional theory \& Tight-binding}
	
	First principles density-functional theory calculations were implemented using the Vienna ab initio Simulation Package (VASP) to calculate the electronic structure of Ca-QFSBLG  \cite{VASP_KRESSE199615}. The Perdew-Burke-Ernzehof (PBE) form of the generalized gradient approximation (GGA) was used to describe electron exchange and correlation \cite{GGA_PBE}. A semi-empirical functional (DFT-D2) was employed to describe van der Waals interactions in the system \cite{DFT_D2_vdWaalsCorr}. The kinetic energy cut-off for the plane-wave basis set was set to 500 eV. We used a 9 $\times$ 9 $\times$ 1 $\Gamma$-centred k-point mesh for sampling the Brillouin zone. The unfolded band structure and Fermi surface were obtained using the KPROJ program based on the k-projection method \cite{KPROJ1,KPROJ2_mgnetic}.
	Tight-binding calculations were performed in Igor Pro Wavemetrics software based on Ref. \cite{AABLgr_TB_Nori2016} for AA-stacked bilayer graphene, and Refs. \cite{GrapheneToGraphite_TB,McCann_Koshino_BLgrTB} for AB-stacked bilayer graphene. Tight-binding calculations are presented along the $\bar{K}-\bar{\Gamma}-\bar{K}$ high-symmetry direction. Parameters used for tight binding calculations were t = 3.05 ($\pm$ 0.05) and $\gamma_1$ = 0.4 for AB-stacked QFSBLG, and t = 2.9 ($\pm$ 0.05) and $\gamma_1$ = 0.4 for AA-stacked Ca-QFSBLG.

	\section{Results and discussion}
	
	\subsection{Experimental results}
	
\begin{figure}[tbh!]
	\centering
	\includegraphics[width=1.0\textwidth]{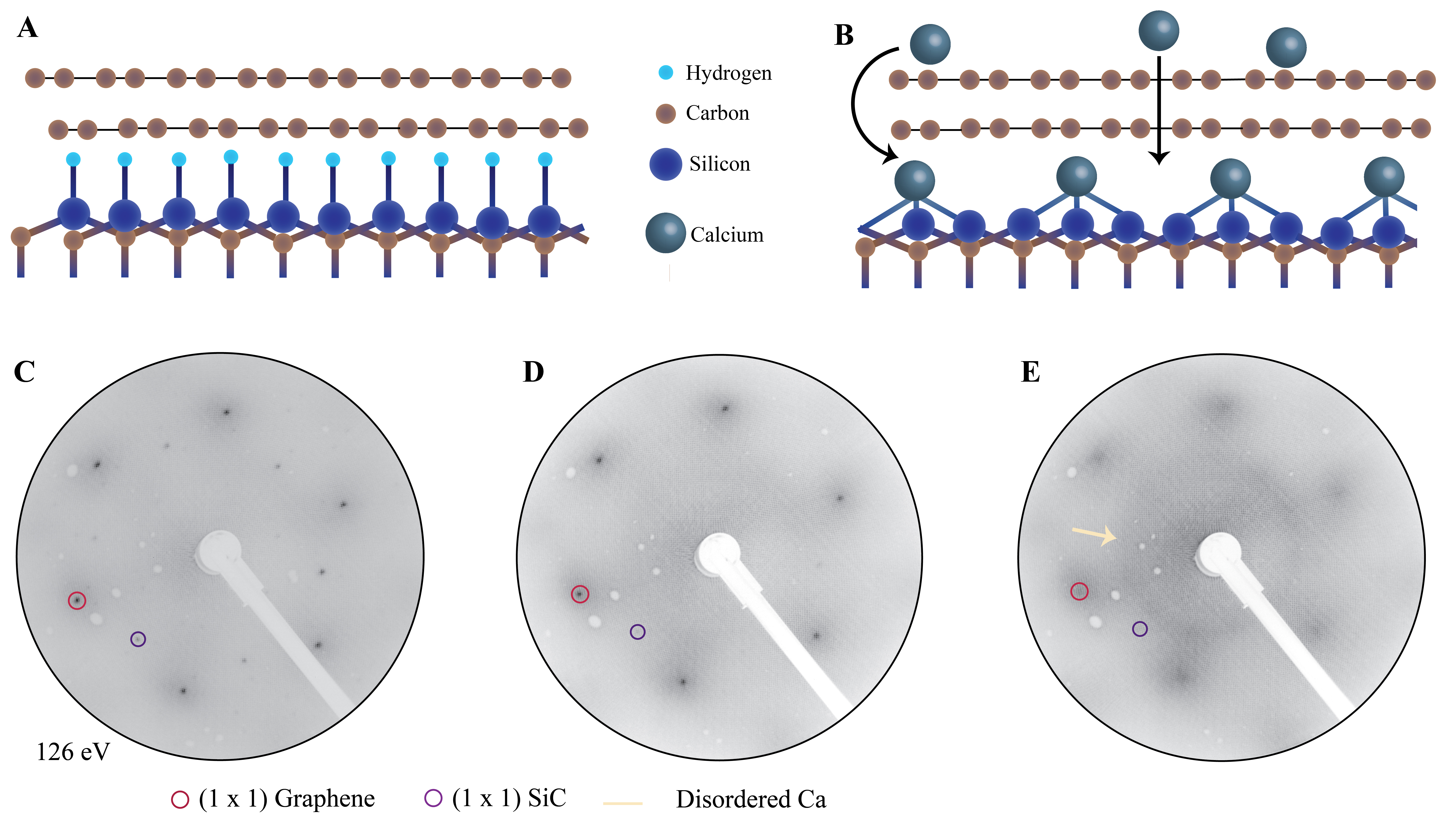}%QFSBLG_Ca_126eV.png}
\caption{Quasi freestanding bilayer graphene (QFSBLG) before and after calcium intercalation. A sketch of QFSBLG before and after intercalation, where hydrogen is replaced by calcium at the SiC interface is shown in \textbf{(A)} and \textbf{(B)}, respectively. \textbf{(C)} LEED image of QFSBLG prior calcium intercalation, \textbf{(D)} following first, and \textbf{(E)} following second calcium intercalation. Red circles indicate the (1 $\times$ 1) graphene spots, purple circles indicate the (1 $\times$ 1) SiC spots. Yellow arrow points to the ring  arising from disordered Ca. All data taken at an incident beam energy of 126 eV and room temperature.}\label{Fig1:LEED}

\end{figure}

QFSBLG samples, Figure \ref{Fig1:LEED}\textbf{A}, prepared as described in Ref.\cite{JimmyXPS} were loaded into the UHV chamber and annealed to remove surface adsorbates, as described in Methods. Following the annealing procedure, LEED data was taken on the clean sample, as shown in Figure \ref{Fig1:LEED}\textbf{C}. LEED data shows typical diffraction pattern of quasi-freestanding graphene, with only (1 $\times$ 1) graphene spots (red circles) and (1 $\times$ 1) SiC spots rotated 30$^\circ$ with respect to graphene (purple circles) visible. Following calcium intercalation (Figure \ref{Fig1:LEED}\textbf{B}), the SiC (1 $\times$ 1) spots are less intense compared to the clean QFSBLG, but no other significant changes can be seen in LEED, as shown in Figure \ref{Fig1:LEED}\textbf{D}. After the second intercalation step, (Figure \ref{Fig1:LEED}\textbf{E}), drastic changes can be observed in the LEED pattern: (1 $\times$ 1) SiC spots are almost completely gone, while (1 $\times$ 1) graphene spots are much weaker and broader. Blurring of the graphene  (1 $\times$ 1) spots suggests additional scattering, likely from calcium atoms accumulating on the surface of the sample in a disordered manner. A new feature can also be observed in the diffraction pattern in Figure \ref{Fig1:LEED}\textbf{E}, a diffuse ring, marked by a yellow arrow, with a radius corresponding to that of a ($\sqrt{3} \times \sqrt{3})$R30$^\circ$ calcium structure. Ring like features observed in LEED usually point towards rotationally disordered system \cite{EMLG_1SiDanglingBondPerUnitCell_JohnRiley2008}, suggesting calcium is not ordered either under graphene or on its surface. This is in contrast to previous LEED data on calcium intercalated graphene on SiC, where a sharp single domain ($\sqrt{3} \times \sqrt{3})$R30$^\circ$ LEED pattern coming from calcium intercalation is observed \cite{C6Ca_CaIntercal_GrapheneThinnestCaC6_Kanetani_PNAS2012, JimmyXPS, Ca_Intercal_SiC_CaLi_HasegawaNew}. One possible explanation for this discrepancy is that in our system, disorder comes from calcium that did not intercalate, but is instead deposited on the surface of the sample where it does not order. Another possibility is that while calcium is replacing hydrogen at the SiC interface, it does so without rotational order, though the reason for this difference in structure is not clear.
	
\begin{figure}[tbh!]
	\centering
	\includegraphics[width=1.0\textwidth]{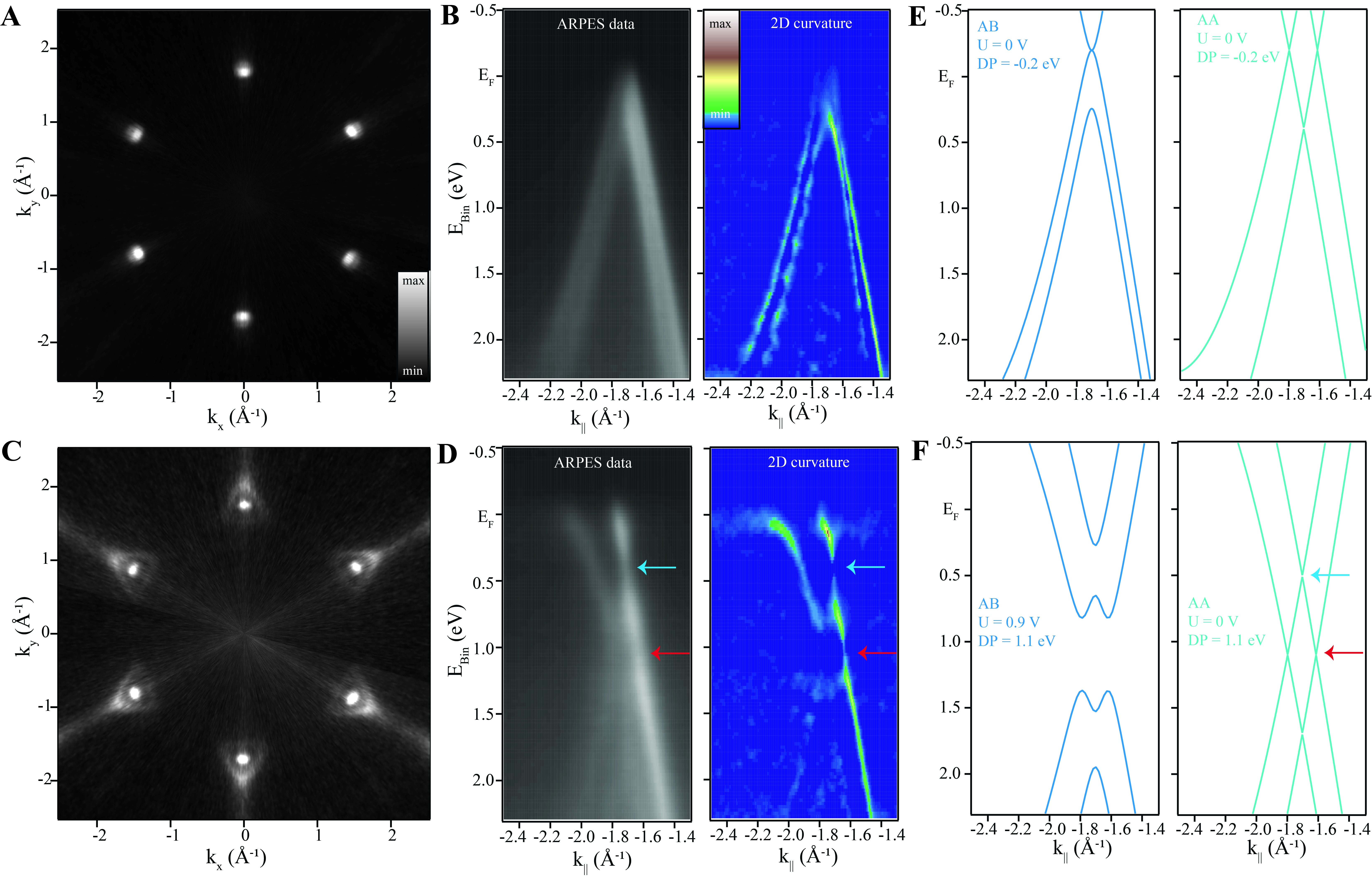}
	\caption{ARPES measurements of clean and calcium intercalated quasi-freestanding bilayer graphene (Ca-QFSBLG) after the first intercalation step. \textbf{(A)} Fermi surface and \textbf{(B)} band structure around the $\bar{K}$ point of QFSBLG, and \textbf{(C)} Fermi surface and \textbf{(D)} band structure around the $\bar{K}$ point of Ca-QFSBLG. In \textbf{(B)} and \textbf{(D)} both raw ARPES spectra (left) and a 2D curvature of ARPES spectra (right) are shown. Following calcium intercalation, a significant change in the electronic structure and doping level can be seen. \textbf{(E)} and \textbf{(F)} show tight-binding calculations for AB- (left) and AA-stacked (right) bilayer graphene before and after calcium intercalation, respectively. Red arrows in \textbf{(D)} and \textbf{(F)} indicate location of band crossing between the top and bottom layer band, and a lack of the band gap for AA-stacked bilayer graphene. Blue arrows in \textbf{(D)} and \textbf{(F)} show the Dirac point in the top graphene layer which appears ungapped.}\label{Fig2:ARPES}
\end{figure}

Following structural characterization by LEED, we proceed with the electronic structure investigation using ARPES \cite{DamascelliARPESrev2004} which allows direct imaging of the electronic bands. Due to an increase in observed disorder for the second intercalation step in LEED, Figure \ref{Fig1:LEED}\textbf{E}, we will only focus on the first intercalation step for the ARPES investigation. Figure \ref{Fig2:ARPES} shows intercalation induced changes in the electronic structure of graphene, as observed by ARPES. Changes are tracked in the energy--momentum cuts taken at the Fermi surface (Figure \ref{Fig2:ARPES}\textbf{A},\textbf{C}) and along the $\bar{K}-\bar{\Gamma}-\bar{ K}$ high-symmetry direction (Figure \ref{Fig2:ARPES}\textbf{B},\textbf{D}). In order to enhance the dispersive features of graphene around the $\bar{ K}$ point, we use a two-dimensional (2D) curvature analysis technique \cite{2D_curvatureMethod}, shown in Figure \ref{Fig2:ARPES}\textbf{B} and \ref{Fig2:ARPES}\textbf{D}, on the right. 
The Fermi surface map Figure \ref{Fig2:ARPES}\textbf{A}, shows the first Brillouin zone of pristine QFSBLG prior calcium intercalation, with six  hole pockets visible at the Brillouin zone boundary. As expected for the case of bilayer graphene \cite{BL_Gr_ARPES_EliScience}, two sets of bands are visible in the energy dispersion data, Figure \ref{Fig2:ARPES}\textbf{B}. Since the samples are p-doped, the Dirac point is located above the Fermi level (0.2 eV, see Supplementary Information), which can be clearly seen in the energy dispersion and the 2D curvature data, Figure \ref{Fig2:ARPES}\textbf{B}. 

Upon calcium intercalation, profound changes can be observed in the electronic structure: two sets of electron pockets can be seen at the Fermi surface (Figure \ref{Fig2:ARPES}\textbf{C}), and system exhibits high levels of n-doping (Figure \ref{Fig2:ARPES}\textbf{D}). In order to better understand the dispersions and changes arising from calcium intercalation, we compare our experimental data to simple tight-binding models for AA- and AB-stacked graphene (Figs. \ref{Fig2:ARPES}\textbf{E},\textbf{F}) \cite{AABLgr_TB_Nori2016,GrapheneToGraphite_TB}. The same model was used to estimate Dirac point position, doping, Fermi wavevector and Fermi velocity. We used the band position data obtained from momentum dispersion curves (MDCs), to refine the tight-binding model, and select the appropriate graphene stacking. More information about this can be found in the Supplementary Information. Prior to calcium intercalation, the system is found to be p-doped, with the Dirac point located at DP = (0.20 $\pm$ 0.02) eV above the Fermi level, and a Fermi wavevector ($k_F$) value of $k_F$~=~(0.057 $\pm$ 0.007) \AA$^{-1}$, corresponding to a hole carrier density of n$_h^{total}$~=~(5.17 $\pm$ 0.08)$\times10^{12}$~cm$^{-2}$, in agreement with literature values for hydrogen intercalated graphene on SiC \cite{Rield_HGR_1stHintercal_PRL2009,H_Gr_SiC_idealGraphene_Starke_PRL2015}.  Following calcium intercalation (Figure \ref{Fig2:ARPES}\textbf{C},\textbf{D}) the sample is transformed into highly n-doped system in which no clear band gap can be seen. This structure is in stark contrast to the C$_6$CaC$_6$ electronic structure, where the ordered calcium phase gives rise to a folding of the $\pi$ bands of graphene, resulting in the folded bands appearing close to the $\bar{\Gamma}$ point \cite{C6Ca_CaIntercal_GrapheneThinnestCaC6_Kanetani_PNAS2012, Ca_Intercal_SiC_CaLi_HasegawaNew, C6CaC6_CaIntercalBLGr_Supercond_Balatski}. In the case of Ca-QFSBLG, no states are observed at the $\bar{\Gamma}$ point, supporting the interpretation that in our case calcium is not ordered in C$_6$CaC$_6$ structure, in agreement with the LEED data (Figure \ref{Fig1:LEED}\textbf{D},\textbf{E}). Absence of the signatures of C$_6$CaC$_6$ structure implies that calcium is not intercalated between the layers, in agreement with recent work by Kotsakidis et al. \cite{JimmyXPS}. Rather, we instead observe a quasi-freestanding bilayer graphene that is n-doped. Nevertheless, there are several discrepancies between our results (Figure \ref{Fig2:ARPES}\textbf{D}) and what is expected from simply n-doped (quasi-freestanding) bilayer graphene \cite{BL_Gr_ARPES_EliScience}. The most obvious discrepancy is a lack of band gap in the system (see red and blue arrow in the Figure \ref{Fig2:ARPES}\textbf{D}), which is expected for the simple case of AB-stacked bilayer graphene \cite{BL_Gr_ARPES_EliScience, Antonija_MgGr_ARPEStoroid}. This is reminiscent of the structure expected for AA-stacked bilayer graphene where a band gap is not expected, and the electronic bands still have massless character as for the case of monolayer graphene \cite{AABLgr_TB_Nori2016}. Taking this into the account, we modelled both AA- and AB-stacked bilayer graphene with tight-binding (Figure \ref{Fig2:ARPES}\textbf{E},\textbf{F}) and DFT (Figure \ref{Fig4:DFT}). From the tight-binding model, the best agreement was obtained for AA-stacked bilayer graphene, where the  inner band Fermi wavevector was $k_F^{inner}$~=~(0.086 $\pm$ 0.007) \AA$^{-1}$, and the outer band was $k_F^{outer}$~=~(0.280 $\pm$ 0.007 )\AA$^{-1}$. These values correspond to a total electron density of n$_e$ = n$_e^{inner}$+n$_e^{outer}$ = (1.37 $\pm$ 0.06) $\times$ 10$^{14}$ cm$^{-2}$, %, with n$_e^{bottom}$ = (2.15 $\pm$ 0.08)$ \times10^{14}$ cm$^{-2}$ and n$_e^{top}$ = (0.14 $\pm$ 0.02) $\times10^{14}$ cm$^{-2}$. %% Move to SI!!!
a two orders of magnitude increase in carrier concentration with respect to the pristine QFSBLG. The aforementioned dramatic increase in the carrier concentration is accompanied with small reduction in Fermi velocity of graphene-- Ca-QFSBLG has Fermi velocity $v_F$ = (0.93 $\pm$ 0.02) $\times10^6$ m/s, while QFSBLG has Fermi velocity $v_F$ = (0.99 $\pm$ 0.02) $\times10^6$ m/s, suggesting that no significant renormalisaiton of Fermi velocity due to transformation of stacking \cite{FermiVelocityRenorm_stacking_Li_Nature2010}, change of the dielectric constant of the graphene's surroundings \cite{FermiVelocityRenorm_dielectric_Lanzara_SciRep2012}, or many-body interactions \cite{FermiVelocityRenorm_manyBody_Geim_NatPhys2011} takes place. Lastly, while the observed increase in the carrier density is significant, it is lower than values observed for the case when calcium goes both to the SiC interface and in between the sheets of bilayer graphene \cite{Ca_Intercal_SiC_CaLi_HasegawaNew}, thus underpinning the notion of intercalation occurring solely at the interface of our system.

\subsection{Theoretical results}

%%% Now in the SI!
%\begin{figure}[tbh!]
%	\centering
%	\includegraphics[width=1.0\textwidth]{img/Sketch_and_DFT.png}
%	\caption{Model of AA- and AB-stacked bilayer graphene with calcium at the SiC interface. Top and side view of calcium intercalated a) AB-stacked and b) AA-stacked bilayer graphene. Distance between the graphene layers, as well as SiC and Ca, and Ca and graphene is labelled in the image.}\label{Fig3:Sketch}
%\end{figure}
We further examine the nature of stacking in Ca-QFSBLG by performing DFT calculations of the electronic structure of AB- and AA-stacked Ca-QFSBLG. We first calculate the calcium intercalation energy based on the Eq.\ref{eq1_CaIntercal} as follows:
\begin{equation}
	E_I = E(SiC/graphene) + E(Ca)  - E(SiC/graphene+Ca),
	\label{eq1_CaIntercal}\end{equation}
where $E_I$ is the intercalation energy, $E(SiC/graphene)$, $E(Ca)$ 
and \\$E(SiC/graphene+Ca)$ are the energy of SiC/graphene heterostructure (SiC covalently bonded with graphene plus a monolayer graphene), atomic energy of calcium in its bulk state and the energy of SiC/graphene system upon calcium intercalation, respectively.
Once calcium is  placed below bilayer graphene, a small difference is found in the formation energy between the AA- and AB-stacked bilayer graphene, as shown in Table \ref{table:structure}, with AB-stacking being slightly favourable. This small difference in formation energy % is likely within the ability for DFT to discriminate, and 
suggests it is plausible that  AA-stacking could indeed be a stable phase in Ca-QFSBLG, similar to what has been observed for the case of lithium intercalation \cite{Li_H_Gr_SiC_Linkopping_SurfSci2012,Li_intercal_EMLG_SiC_Johansson_2016}. 
As both structures appear energetically stable, and are close in the formation energy, DFT band structure calculations were performed for AA- and AB-stacking in order to determine which  structure fits experimental ARPES data better.

\begin{table}[tbh!]
	\centering
	\begin{tabular}{c c c } 
		\hline\hline
		Stacking type & DFT-D2 energy [eV] & DFT-D3 energy [eV] \\ [0.5ex] 
		\hline\hline
		AA & 0.94 & 1.08\\ 
		\hline
		AB & 0.98 & 1.09 \\
		
		\hline\hline
	\end{tabular}
	\caption{Calculation for Intercalation energy, AA vs. AB, using different van der Waals corrections.}
	\label{table:structure}
\end{table}

\begin{figure}[tbh!]
	\centering
	\includegraphics[width=1.0\textwidth]{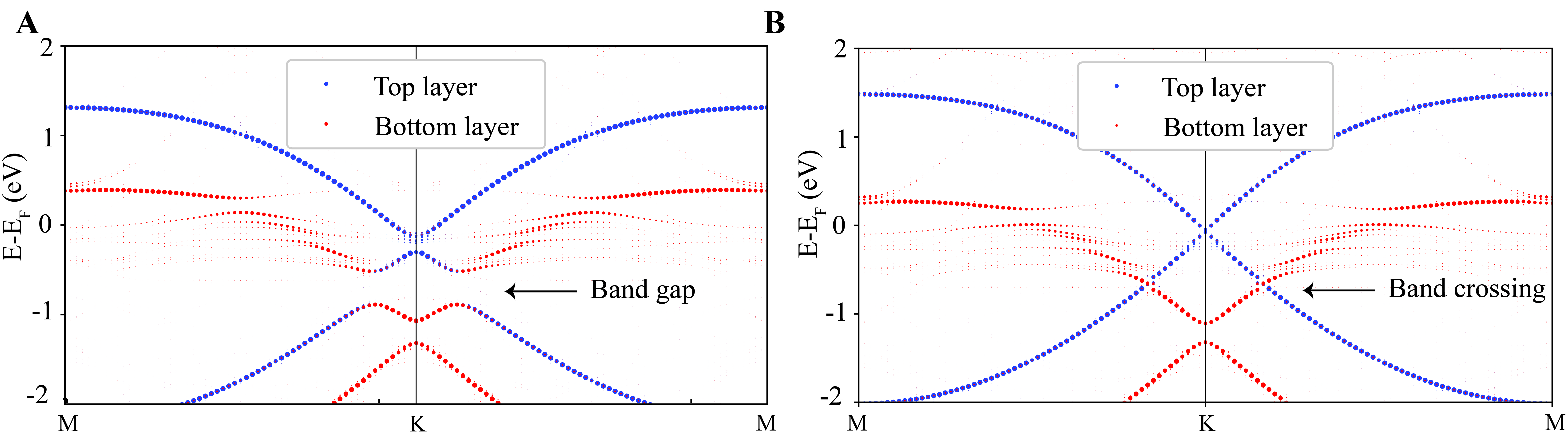}
	\caption{DFT calculations for calcium intercalated QFSBLG. The unfolded band structure of calcium intercalated bilayer graphene for the case of AB- and AA-stacking is shown in \textbf{(A)} and \textbf{(B)}, respectively. The contribution from the top graphene layer is shown in blue, and from the bottom layer in red. Arrows indicate a band gap for the case of AB-stacking, \textbf{(A)}, and a band crossing for the case of AA-stacking in Ca-QFSBLG,  \textbf{(B)}.}\label{Fig4:DFT}
\end{figure}
%\clearpage

The unfolded band structure of AA- and AB-stacked Ca-QFSBLG is shown in Figure \ref{Fig4:DFT}. For the case of AB-stacked Ca-QFSBLG (Figure \ref{Fig4:DFT}\textbf{A}), a large band gap, approximately 0.38 eV in size, is found between the top of the valence band and bottom of a conduction band, situated 0.51 eV below the Fermi level. This structure is similar to the one observed for magnesium intercalated graphene on SiC \cite{Antonija_MgGr_ARPEStoroid}, where a band gap of 0.35 eV was observed. In the case of AA-stacked bilayer graphene (Figure \ref{Fig4:DFT}\textbf{B}), the structure is markedly different, and no band gap is found between the top of the valence band and bottom of a conduction band. Instead, a smaller 0.2 eV gap is identified 1.05 eV below the Fermi level located only in the bottom graphene layer. In contrast, the top graphene layer is gapless, and nearly indistinguishable from the pristine monolayer graphene. The latter structure is in good agreement with the experimental data shown in Figure \ref{Fig2:ARPES}, particularly in the region where the top- and bottom-layer derived bands cross at finite momentum (red arrows in Figs. \ref{Fig2:ARPES}\textbf{D} and \ref{Fig2:ARPES}\textbf{F}) and the Dirac point of the top graphene layer (blue arrows in Figs. \ref{Fig2:ARPES}\textbf{D} and \ref{Fig2:ARPES}\textbf{F}), in agreement with our experimental ARPES results which show Ca-QFSBLG as AA-stacked.

\section{Conclusion}
Calcium intercalation was successfully achieved in quasi-freestanding bilayer graphene on hydrogenated SiC, resulting in significant changes to the system. Upon calcium intercalation, calcium replaced hydrogen at the SiC interface, leading to a switch from p-type doping to n-type doping. This transition was accompanied with almost two orders of magnitude change in the carrier concentration, going from n$_h^{total}$~=~5.17~$\times10^{12}$~cm$^{-2}$ to n$_e^{total}$ = 1.37 $\times$ 10$^{14}$ cm$^{-2}$, while retaining the quasi-freestanding nature, and exhibiting minimal change of the Fermi velocity. Structurally, the intercalation process resulted in a transformation from AB-stacked to AA-stacked bilayer graphene, a shift facilitated by a small difference in the formation energy between the two stacking types. As a result, the electronic band structure is significantly altered. The top layer of graphene is nearly indistinguishable from a pristine monolayer, retaining the ungapped Dirac point. While previous reports show indirect evidence of inhomogeneous AA- and AB-stacked regions in epitaxial graphene following lithium intercalation and de-intercalation, to our knowledge, this is the first report of the preparation of large-area, uniform quasi-freestanding AA-stacked graphene following calcium intercalation. This opens the door to further study of the properties of this distinct new graphene system with its unique band structure.   

\section*{Acknowledgement}
This work was supported by the Australian Research Council under awards DP150103837, DP200101345 and FL120100038. This research was undertaken on the Soft X-ray spectroscopy beamline at the Australian Synchrotron, part of ANSTO. J.C.K. gratefully acknowledges support from the Australian Government Research Training Program, and the Monash Centre for Atomically Thin Materials. Y.Y. and N.M. gratefully acknowledge the support from the Australian Research Council (CE17010039) and the computational support from the National Computing Infrastructure and Pawsey Supercomputing Facilities. D.K.G., R.L.M-W., and K.M.D. acknowledge support by core programs at the U.S. Naval Research Laboratory funded by the Office of Naval Research.

\bibliographystyle{ieeetr}
\bibliography{main}

\end{document}